\documentclass[letterpaper]{article} 
\usepackage{aaai24}  
\usepackage{times}  
\usepackage{helvet}  
\usepackage{courier}  
\usepackage[hyphens]{url}  
\usepackage{graphicx} 
\usepackage{natbib}  
\usepackage{caption} 
\usepackage{algorithm}
\usepackage{algorithmic}
\usepackage{url}
\usepackage{booktabs}
\usepackage[table]{xcolor}
\definecolor{lightgray}{gray}{0.9}

\usepackage{newfloat}
\usepackage{listings}
\DeclareCaptionStyle{ruled}{labelfont=normalfont,labelsep=colon,strut=off} 
\floatstyle{ruled}
\newfloat{listing}{tb}{lst}{}
\floatname{listing}{Listing}
\newenvironment{quoteitalicized}
    {\begin{quote}}
    {\end{quote}}
\newcommand{\quotes}[2]{\begin{quoteitalicized}\textit{#1} -{#2}\end{quoteitalicized}}

\nocopyright

\title{AI Governance in Higher Education:\\Case Studies of Guidance at Big Ten Universities}
\author{
    Chuhao Wu, He Zhang, John M. Carroll
}
\affiliations{

    College of Information Science and Technology\\
    The Pennsylvania State University\\
}

\begin{document}

\maketitle

\begin{abstract}
Generative AI has drawn significant attention from stakeholders in higher education. As it introduces new opportunities for personalized learning and tutoring support, it simultaneously poses challenges to academic integrity and leads to ethical issues. Consequently, governing responsible AI usage within higher education institutions (HEIs) becomes increasingly important. Leading universities have already published guidelines on Generative AI, with most attempting to embrace this technology responsibly. This study provides a new perspective by focusing on strategies for responsible AI governance as demonstrated in these guidelines. Through a case study of 14 prestigious universities in the United States, we identified the multi-unit governance of AI, the role-specific governance of AI, and the academic characteristics of AI governance from their AI guidelines. The strengths and potential limitations of these strategies and characteristics are discussed. The findings offer practical implications for guiding responsible AI usage in HEIs and beyond.

\end{abstract}

 \section{Introduction}
AI applications in education (AIEd) have been at the forefront of discussions among education stakeholders, particularly in light of the advancements in Generative AI (GenAI). Undoubtedly, the continually improving performance of AI holds immense potential for enhancing educational experiences, enabling personalized learning, and automating administrative tasks \cite{baidoo2023education,chen2020artificial}. However, alongside these benefits, there are legitimate concerns about the potential negative impact of AI, especially GenAI. The fact that AI is trained with existing work and its ability to quickly generate content that may be unauthorized imitations impose fundamental challenges to the regulation of plagiarism and academic integrity \cite{dehouche2021plagiarism,smits2022generative}. These challenges will potentially impact both faculty and students in their work and study. Specifically, for educators, the need to learn about AI and address it in teaching brings extra requirements for their knowledge and skills \cite{wu2023integrating}. For students, the prevalence of AI also raises concerns about ethical issues, personal development, career prospects, and societal values \cite{chan2023students}. 

Higher education institutions (HEIs) play a crucial role in technology innovation and its diffusion through creating knowledge, providing talent, and translating research into innovations \cite{gunasekara2006reframing}. In the development and implementation of norms for AI governance, academic institutions responded by establishing centers for the study of AI governance, engaging in basic research, and influencing AI norms through academic expertise and collaborations \cite{chinen2023ai,mainzer2022responsible}. While HEIs contribute to AI governance at a societal level, technology diffusion in universities itself can also be a complex process involving issues of power, legitimacy, and identity \cite{smith2000higher}. In the case of GenAI, due to the complex implications of encouraging or discouraging such an important technology, universities are urged to take action and devise strategies for guiding and regulating its usage among students, faculty, and staff \cite{chan2023comprehensive}. Despite significant uncertainty about the pros and cons of incorporating GenAI in education, many universities have published policies and guidelines, aiming to promote responsible and beneficial AI usage. For example, the early review conducted by \citet{moorhouse2023generative} found that 23 out of the 50 top-ranking universities have developed publicly available guidelines by June 2023, which addressed the potential influence of GenAI on three main areas: academic integrity, advice on assessment design, and communicating with students. However, the authors pointed out that as these guidelines have been developed rapidly, it is likely that many of the suggestions have not been sufficiently tested. Similarly, \citet{adams2015using} examined documents produced by 116 high research activity (R1) universities in the US, concluding that the majority of universities encourage the use of GenAI yet in a way that presents potential burdens for faculty and students and without much regard for ethical concerns.

The current study will contribute to this line of research by extending the focus beyond teaching and learning implications and analyzing the guidelines through the lens of AI governance. Prior studies on AI guidelines have provided an overview of how many HEIs are embracing GenAI usage and publishing AI guidelines, and what main topics are discussed. This study aims to further analyze the structure of AI guidelines from the organizational level and a community perspective. Specifically, we aim to illustrate the approaches HEIs take to promote responsible AI usage by answering the following research questions:
\begin{enumerate}
    \item What strategies of AI governance are demonstrated by these guidelines?
    \item What are the characteristics of the guidance provided to the university community?
\end{enumerate}

\section{Background}
\subsection{Responsible AI Governance}
As AI technology continues to advance and to be widely adopted, the governance of its development and implementation has become a key issue that is widely concerned and discussed by various stakeholders~\cite{10.1145/3278721.3278731,thiebes2021trustworthy,10.1145/3514094.3534146}. AI governance refers to the frameworks, policies, and practices designed to ensure that AI systems are developed, deployed, and managed in a way that aligns with ethical principles, legal requirements, and societal values. It encompasses the establishment of guidelines for responsible AI use, mechanisms for accountability, and processes for managing risks associated with AI technologies ~\cite{felzmann2020towards,balasubramaniam2023transparency,rismani2023does,10.1145/3461702.3462564}. Organizations establish ethical guidelines for AI usage by engaging in a multifaceted process that involves the identification of core ethical principles, the development of practical frameworks, and the implementation of governance models that ensure adherence to these guidelines throughout the AI system's lifecycle~\cite{janssen2020data,georgieva2022ai,10.1145/3514094.3534195}. The process often begins with the recognition of the need for ethical AI principles that sustain human values and rights~\cite{diaz2023connecting}, acknowledging the complex and opaque nature of AI systems and their potential to impact fairness, accountability, and transparency~\cite{akinrinola2024navigating,diakopoulos2020accountability}. This recognition is supported by international and organizational efforts to publish principles of ethical AI, which aim to outline values and abstract requirements for AI development and deployment \cite{rees2023all}. However, the effectiveness of these principles is contingent upon their translation into measurable and actionable guidelines that can be practically applied \cite{fellander2022achieving}. 

To address the gap between ethical principles and their application, organizations are adopting frameworks like the hourglass model of organizational AI governance. This model emphasizes the need for governance at environmental, organizational, and AI system levels, connecting ethical principles to the AI system lifecycle to ensure comprehensive governance \cite{huriye2023ethics}. These collective efforts underline the importance of a structured approach to AI governance, ensuring that AI technologies are developed and deployed in ways that uphold ethical standards and promote social good~\cite{smuha2019eu}. HEIs also make prominent contributions to the process as research efforts are critical in assessing the impact of intelligent applications, minimizing harm, and promoting well-being and social good \cite{safdar2020ethical}. In addition, a comprehensive understanding of the impact of disruptive AI technologies on education and the development of frameworks to critical for society to make informed decisions about the use of generative AI tools \cite{khan2024ethics}.  This study aims to bridge the gap between theoretical principles and practical application, offering a detailed analysis of whether and how HEIs implement responsible AI governance frameworks.


\subsection{Technology Diffusion in HEIs}

Technology diffusion refers to the process by which new technologies are adopted and spread across different regions, sectors, and social groups, during which technology is transferred, implemented, and utilized, leading to widespread acceptance and integration into everyday practices \cite{rogers2014diffusion}. Technology diffusion in universities is a complicated process, which often faces multifaceted challenges rooted in both organizational and cultural aspects of HEIs~\cite{hawawini2011internationalization,liu2020understanding}. One primary challenge is the rapid and massive development of technology, which requires continuous adaptation and innovation from HEIs to maintain their educational and academic quality~\cite{christensen2011innovative}. This adaptation is contingent upon the effective diffusion of innovation models that consider the profile of human resources, technological conditions, organizational policies, documentation, and environmental dynamics \cite{ramdhani2021diffusion}. Additionally,  the adoption of new information and communication technology is hindered by issues such as incompatible technology with faculty's traditional teaching practices, inadequate faculty support, and insufficient plans for implementation \cite{dintoe2019technology}. Instructors face challenges in staying updated with the changing uses of technology in the classroom, necessitating training, funding, and alignment of perceptions to facilitate technology adoption \cite{baadel2017technology}. The integration of technology in teaching and learning processes is influenced by technological, pedagogical, and organizational dimensions, with contextual factors playing a determinant role \cite{rodriguez2020digital}. 

Regarding the adoption of AI in HEIs, there is still a lack of sound evidence on the pedagogical impact of AI technologies, which raises questions about their ability to improve learning outcomes or facilitate effective pedagogical changes \cite{o2023artificial}. Additionally, the ethical concerns associated with biased algorithms, which could adversely affect students if used in admissions or grading processes, represent a significant technological and ethical barrier \cite{slimi2023navigating}. Educators' perspectives and the social, psychological, and cultural factors influencing their trust and adoption of educational technology also play a role in the slow uptake of AI tools \cite{kizilcec2024advance}. The barriers to digital technology integration, including technophobia and the absence of planning, directly impact the adoption of AI in university teaching \cite{mercader2020explanatory}. Therefore, to effectively govern the implementation of AI tools, it is crucial to integrate robust communication practices. AI governance in HEIs involves not just creating policies but also ensuring they are clearly communicated and understood at all levels. This includes identifying key stakeholders responsible for crafting and delivering these messages, such as senior leaders or governance committees. Effective communication from authoritative figures emphasizes the importance of these guidelines and fosters a culture of ethical compliance. Thus, a connection between governance and AI guidelines is established through the need for clear, consistent, and authoritative communications. Through the analysis of public guidelines on GenAI, we aim to shed light on how HEIs could actively adopt AI technology and how it manages the potential harms through communications with its community. These communications, in the form of policy documentation and webpages, facilitate understanding, transparency, and collaboration between the institution and its stakeholders and effectively convey how the institution is integrating AI technology and addressing any associated risks.  

\section{Methods}

\begin{table*}[h]
\centering
\caption{The background information of the Big Ten universities}
\label{tab:bigten}
\resizebox{\linewidth}{!}{%
\begin{tabular}{lllll}
\hline
\textbf{ID} & \textbf{Institution}                    & \textbf{Location}                      & \textbf{Type}            & \textbf{Enrollment} \\ \hline
U1           & University of Iowa                      & Iowa City, Iowa                        & Public                   & 30015               \\
U2           & University   of Wisconsin–Madison       & Madison,   Wisconsin                   & Public   (land-grant)    & 51528               \\
U3           & University of Maryland, College   Park  & College Park, Maryland                 & Public (land-grant)      & 40792               \\
U4           & Michigan   State University             & East   Lansing, Michigan               & Public   (land-grant)    & 50023               \\
U5           & Pennsylvania State University           & University Park, Pennsylvania          & Public (land-grant)      & 50028               \\
U6           & Indiana   University Bloomington        & Bloomington,   Indiana                 & Public                   & 47005               \\
U7           & University of Michigan                  & Ann Arbor, Michigan                    & Public                   & 51225               \\
U8           & University of Minnesota, Twin Cities  & Minneapolis-St.  Paul, Minnesota      & Public   (land-grant)    & 54955               \\
U9           & Rutgers University-New Brunswick        & New  Brunswick–Piscataway, New Jersey & Public (land-grant)      & 50637               \\
U10          & Purdue University                     & West  Lafayette, Indiana              & Public   (land-grant)    & 45869               \\
U11          & University of Nebraska–Lincoln          & Lincoln, Nebraska                      & Public (land-grant)      & 23805               \\
U12          & Northwestern  University               & Evanston,  Illinois                   & Private   not-for-profit & 23161               \\
U13          & Ohio State University                   & Columbus, Ohio                         & Public (land-grant)      & 60540               \\
U14          & University of Illinois Urbana–Champaign & Urbana-Champaign, Illinois             & Public (land-grant)      & 56916               \\ \hline
\end{tabular}%
}
\end{table*}

\subsection{The Big Ten Universities}
The Big Ten Conference is the oldest collegiate athletic conference in the United States. While historically celebrated for its athletic prowess, its member institutions are also major research universities with large financial endowments and strong academic reputations. Established in 1896, the Big Ten has grown to include fourteen universities spread across the Midwest and Northeast. These institutions are known for their research and academic excellence, contributing significantly to advancements in science, technology, engineering, and mathematics (STEM) education. The list of the 14 universities, their locations, and enrollments are presented in Table~\ref{tab:bigten} (On August 2, 2024, the conference expanded to 18 member institutions and 2 affiliate institutions. The additional 4 members and affiliate members were not included in this study).

We chose this list of institutions for the case study due to the following reasons. First, the Big Ten is known for its research contributions and scholarly output. They often have substantial funding and resources dedicated to technological advancement, including AI, making them influential in shaping research agendas and policies related to AI nationally and internationally. Studying their guidance on AI can provide insights into how leading research institutions envision the future of AI governance and ethical considerations. The Big Ten also shares similar profiles as large, research-intensive institutions with significant resources and academic influence. This homogeneity provides a level of consistency in comparing AI guidelines, making it easier to identify patterns and common themes within a similar institutional context. Focusing on this list limits the sample to a manageable number of institutions, which facilitates a more in-depth and detailed analysis of each university’s AI guidelines and policies. Previous research has also utilized this sample to study campus recreation \cite{wilson2020big}, ethnicity diversity \cite{bennett2002enhancing}, sexual harassment \cite{clair1993bureaucratization}, and transgender policy \cite{dirks2016transgender} in HEIs. Therefore, focusing on the Big Ten allows us to identify insights on AI governance that are interpretable and meaningful to other HEIs and society.

\subsection{Data Collection}
The data collection was conducted in March 2024. We define AI guidance as guidelines regarding AI usage published officially at the university level. Therefore, guidelines published by a specific college or a branch campus are not included in the analysis. To identify AI guidelines for each university, the primary author visited the 14 universities' official websites and conducted manual searches. The keywords `AI', `Generative AI', `Guidance', `Guideline', and `Policy' are used in combination to conduct the search. Additionally, we used the keywords appended with the university name in Google Search and inspected the results returned on the first page to ensure comprehensive coverage in our search. After the initial round of data collection, another researcher repeated the search process independently to validate the results and check for missing information. All of the 14 universities have official guidelines related to AI. Only publicly accessible documents or websites were extracted for further analysis. The data collection and analysis process was reviewed and approved by the IRB office in the authors' home institution.

\subsection{Data Analysis}
Mind mapping technique \cite{wheeldon2009framing} and thematic analysis \cite{drisko2016content} were employed to analyze the guidelines. First, the primary author read through all the collected data and created mind maps using the university as the base unit. Each mind map represents the structure of AI guidelines published by the university, including the specific unit that publishes the guideline and the organization of its content. The mind maps enable an overview of how guidelines are organized by each university, revealing commonalities and differences in AI governance. The initial mind maps were then reviewed collectively by all researchers to ensure the accurate representation of the original data. In the next round of data analysis, the primary author conducted a thematic analysis of the mind maps, systematically coding and categorizing the mind map nodes while referencing the original guideline text to maintain contextual accuracy. Specifically, topics and segments that represented the strategies and characteristics of the AI guidance were identified, categorized, and linked to broader themes within the research. This thematic analysis not only highlighted the core elements of AI guidance but also allowed for the identification of patterns, relationships, and emerging trends within the data, thereby providing a deeper understanding of the subject matter. The coding was then shared and reviewed by all researchers, who discussed the consistency in coding and resolved minor discrepancies. The codes were grouped into three major themes: the multi-unit governance of AI, the role-specific governance of AI, and the academic characteristics of AI governance. After the researchers completed the process, they collectively translated the themes and sub-themes into narratives. The mind maps of the 14 universities can be found in the supplementary materials.

\section{Results}
Instead of specific rules or suggestions for AI usage, our analysis focuses on the strategies taken by universities to guide AI usage and the common characteristics observed in these AI guidelines. The findings and insights are detailed in the following subsections.

\subsection{Multi-unit Governance of AI}
\begin{table*}[ht]
\centering
\caption{Units that publish AI guidelines on a university level}
\label{tab:unit}
\rowcolors{1}{}{lightgray}
\begin{tabular}{lllllllllllllll}
\hline
ID       & U1 & U2 & U3 & U4 & U5 & U6 & U7 & U8 & U9 & U10 & U11 & U12 & U13 & U14 \\ \hline
Information Technology       & $\bullet$ & $\bullet$ &   & $\bullet$ &   & $\bullet$ &  $\bullet$  & $\bullet$  &   &    &    & $\bullet$  &    &    \\
Teaching \& Learning       & $\bullet$ & $\bullet$ & $\bullet$ & $\bullet$ &   & $\bullet$ &   & $\bullet$ & $\bullet$ & $\bullet$  & $\bullet$  & $\bullet$  & $\bullet$  & $\bullet$  \\
President \& Provost   & $\bullet$ &   & $\bullet$ &   &   &   &   & $\bullet$ &   & $\bullet$  &    &    &    &    \\
University Libraries   &   & $\bullet$  &   &   &   &   & $\bullet$  & $\bullet$ & $\bullet$ &    &    & $\bullet$  &    &    \\
AI Center &   &   &   &   & $\bullet$ &   & $\bullet$ &   &   &    &    & $\bullet$  &    &    \\
Additional units     &   & $\bullet$ &   & $\bullet$  &   & $\bullet$ &   & $\bullet$  &   &    &    &    &    &    \\ \hline
\end{tabular}
\end{table*}

The adoption and management of new technologies in HEIs can be complex due to their diverse constituencies, including faculty, students, and staff, each with distinct needs and priorities. In the case of the Big Ten, multiple units have been involved in publishing AI guidelines for the university community, as shown in Table~\ref{tab:unit}. While the advice offered by different units can still overlap and reference each other, some key differences entail the unique role of each unit in the organizational management and AI governance in HEIs.

\subsubsection{Information Technology.} Information Technology (IT) and similar departments in HEIs play a crucial role in managing network services, supporting various technologies and platforms, and resolving software and hardware issues encountered by students and staff \cite{volk2018landscape}. Six universities from the Big Ten have issued AI guidelines through their IT departments. These guidelines mainly focused on security and privacy risks associated with Gen AI tools and how to avoid them. Specifically, these risks are managed by IT through statements related to the following topics:

\paragraph{1. Data-sharing policy.} IT established a specific policy for data usage as the community interacts with AI tools. These statements guide people to input limited types of data into AI tools and prohibit the sharing of institutional data and other sensitive information, which is also in accordance with the university's existing data classification system: 

\quotes{Do not share institutional data with this tool. Providing any personally identifiable information or university internal information, such as development code for systems hosting institutional data, is a violation of IU policy.}{U6}

The data classification can be found on other pages of the IT website, which usually categorize data into four levels: Public (Low Sensitivity), University Internal (Moderate Sensitivity), Sensitive/Restricted (High Sensitivity), and Restricted/Critical (Very High Sensitivity). Among the four levels, the Public data is considered safe to share without the need for further IT review. By recapitulating this data classification in AI guidelines, IT helps university users avoid undesired information leakage while using AI.

\paragraph{2. Enterprise agreement.} IT also provides information on AI tools that are considered more secure through an enterprise contract or agreement with the university. These tools are approved for data interaction classified up to and including the \textit{University Internal} level. The most common tool with such an agreement is Microsoft Co-Pilot, due to part of the university's existing licensing with Microsoft services such as Office 365. Therefore, despite the agreement, IT still seeks to enhance risk avoidance through further qualifications of AI tools:

\quotes{Since Microsoft released this tool as part of existing licensing, it has not yet gone through the formal review process that is part of our standard procurement process. As with any such tool, caution is advised.}{U4}

As this statement shows, IT has a formal review process that ensures a tool or software meets the university's security requirements. The guidelines welcome faculty to contact IT to assess security attributes for a given implementation of AI.

\begin{figure}[ht]
\includegraphics[width=\linewidth]{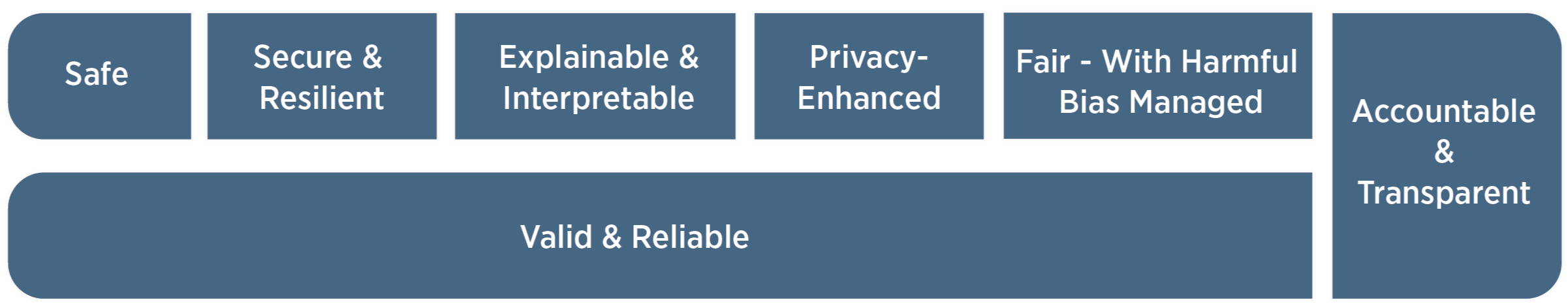}
\caption{Characteristics of trustworthy AI systems defined by NIST: Valid \& Reliable is a necessary condition of trustworthiness and is shown as the base for other trustworthiness characteristics. Accountable \& Transparent is shown as a vertical box because it relates to all other characteristics \cite{tabassi2023artificial}.}
\label{fig:trust_ai}
\end{figure}

\paragraph{3. Trustworthy AI.} While it's impossible for IT to comment on all AI tools available on the market, it has encouraged the community to use ``Trustworthy AI'' tools. The characteristics of ``Trustworthy AI'' are described using the existing National Institute of Standards and Technology's AI Risk Management Framework (see Figure~\ref{fig:trust_ai}):

\quotes{UW Madison faculty, staff, students and affiliates can help protect themselves and others by choosing tools and services that exhibit the NIST’s characteristics of trustworthy AI.}{U2}

By referencing NIST's framework, IT promotes responsible AI usage in compliance with national standards.

\subsubsection{Teaching \& Learning.} These units are university departments committed to providing various services and solutions to enhance teaching and learning practices. With the grand objective of supporting teaching and learning, some of these departments also emphasize the role of technology in empowering students and instructors:

\quotes{The ITS Office of Teaching, Learning, and Technology provides expertise, tools, and services to optimize teaching and learning through learning sciences research, ICON, teaching and learning data, and advanced classroom and instructional technology.}{U1}

Naturally, these units aim to provide guidelines for faculty to understand GenAI and how it can impact their teaching practices. As teaching is a core activity for universities,  these guidelines involve a diverse list of implications of AI, covering the benefits of personalized learning, ethical issues with education equity, and long-term impact on higher education. We summarized three common approaches the guidelines take to shape responsible AI behaviors among faculty:

\paragraph{1. Emphasize the Need to Learn about GenAI.} These guidelines provide a useful introduction to GenAI and related concepts in the format of text or video. Faculty are encouraged to learn more about the benefits and limitations of GenAI not only by reading and viewing these resources but also by experimenting with AI tools :

\quotes{Learn what AI tools can and cannot do by reading up on these tools and experimenting with them before incorporating an AI tool into a class activity or restricting its use.}{U9}

`Frequently asked questions' is a commonly used format to describe how AI could be appropriately used by faculty, e.g., ``What are some examples of assessments that incorporate AI tools?''. While Teaching \& Learning attempts to help faculty learn the implications of AI in a broad range of aspects, U13 also points out that ``much of what we know about generative AI applications and AI language models will shift over time''. Therefore, these guidelines emphasize that faculty need to take active actions to become more knowledgeable on AI technology.

\paragraph{2. Example-based recommendations.} In addition to learning about GenAI, these AI guidelines make recommendations for faculty to modify their teaching practices to address potential issues brought by AI. These recommendations cover the full cycle of education delivery, including course preparation, class management, assignment design, and grading methods, with examples to follow. The examples aim to help faculty integrate AI in teaching and regulate AI usage among students. Importantly, they also highlight the fact that efforts are needed to ensure the essential objectives of teaching are not disrupted by the existence of AI tools:

\quotes{Consider developing assignments that require students to use higher-order thinking, connect concepts to specific personal experiences, cite class readings and discussions, and make innovative connections. These types of prompts are more difficult for students to answer using AI tools.}{U11}

\paragraph{3. Responsibilities for guiding student usage.} Teaching \& Learning mostly provide guidance for faculty and instructors. By doing so, it also delegates the responsibilities of guiding students to faculty. Specifically, the guidelines recommend instructors to ``develop clear policies for each course'' that specify whether and when AI usage is allowed, communicate expectations with students in ``inclusive, student-centered language'', and ``include time for ethics discussions'' that help students to not only use AI properly but also learn the implications of using AI. Another important responsibility for faculty is to protect academic integrity as GenAI is generally considered a threat to academic integrity. Instead of simply restricting or punishing the use of GenAI, the guidelines encourage faculty to foster a positive mindset and emphasize the values of academic integrity for students. For example:

\quotes{Share your perspectives on how you think these tools can help or hinder their learning, and why you value academic integrity. We suggest focusing on the benefit to students and their learning, and not potential negative consequences to their grade.}{U10}

\subsubsection{President \& Provost.}
The Office of the President or Office of the Provost typically plays a central role in academic administration, serving as the chief academic officer of an institution. However, in terms of AI governance, the role of the President \& Provost seems ambiguously defined, with guidance content overlapping with several other units. For example, in U3, it provides high-level summary guidance for students, faculty, and staff while Teaching \& Learning provides more comprehensive guidance for faculty. Similarly, in U1, President \& Provost mainly recapitulates issues addressed by other units and serve to raise people's awareness of available resources:

\quotes{The Office of Teaching, Learning, and Technology and the Center for Teaching have put together an AI Tools and Teaching webpage that includes sample policy language for the use of AI tools in a variety of contexts that you can incorporate into your syllabus.}{U1}

\subsubsection{University Libraries.}
Libraries often provide a wide range of support to the university community, including research, learning, and engagement. The guidance published by libraries mainly targets scholarly activities, which can include teaching activities but focus more on research and academic publishing. The content provides instructions for scholars to improve their productivity with AI but also raises awareness of the complications involving accuracy, bias, academic integrity, and intellectual property.

\quotes{Although it's not advised to use generative AI directly to find sources on your topic, AI chatbots may be helpful for some parts of your research process.}{U2}

The statement above indicates an awareness of both the potential risks and the improvement of productivity by using AI in research. Therefore, AI guidance by libraries features both recommendations for AI tools and ethical considerations for using them, e.g.,  citing AI-generated content to address academic integrity concerns:

\quotes{Before including AI-generated content in a project you intend to get published, check publisher policies regarding permissible use and attribution. Below are some examples of publisher policies regarding the use of AI.}{U9}

Similar to the Trustworthy AI framework, the reference to example policies from well-known publishers shows that institutions try to shape responsible AI behaviors that are in compliance with existing policies and standards from well-known organizations.

\subsubsection{AI Center.}
\begin{figure}[ht]
\includegraphics[width=\linewidth]{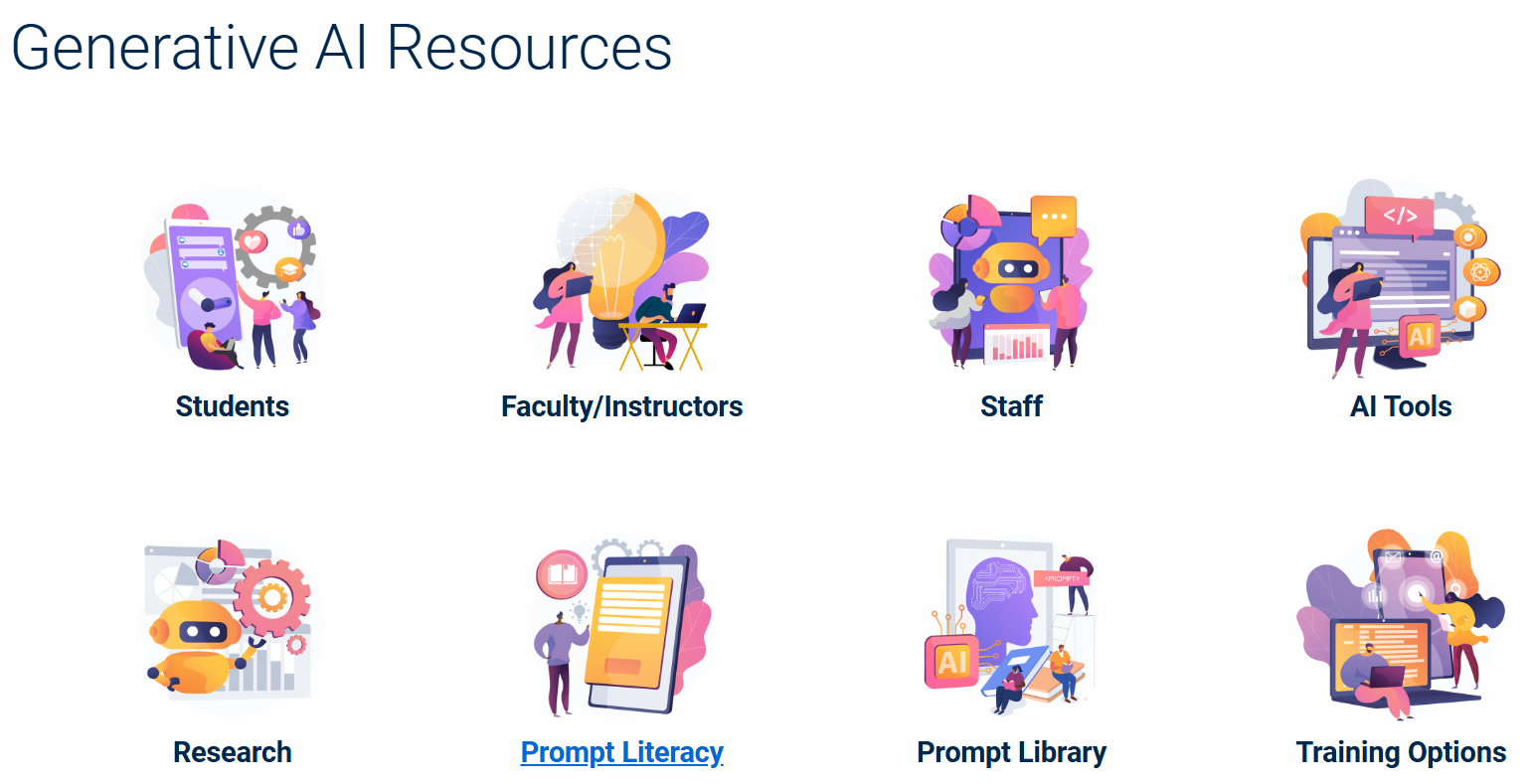}
\caption{GenAI guidance published by AI Center at U7, categorized by student, faculty, and staff}
\label{fig:um_ai}
\end{figure}

AI Centers refer to units specifically devoted to AI-related issues in the university. In the case of U12, the AI Center is a website focused generally on AI-related research and education, broadcasting AI activities on campus. It also provides AI guidance that resembles what is offered by Teaching \& Learning in other institutions:  
\quotes{This website is intended to acquaint you with GAI and to give you some suggestions for its use in the classroom.}{U12}
In contrast, the AI Center at U7 is a website dedicated to GenAI but offers comprehensive guidelines covering students, faculty, and staff (see Figure~\ref{fig:um_ai}). Beyond the impact on teaching and the classroom, it addresses enterprise agreements, data privacy, and research usage, topics typically covered by other units.
It is possible for an institution to have more than one AI Center. In the case of U5, the `AI, Pedagogy, and Academic Integrity' website focuses on the impact of GenAI, with content similar to that of Teaching \& Learning, mainly discussing the impact of AI in the classroom. In addition, the `AI Hub' addresses all AI-related research and education news, and its guidance covers accessibility and data security considerations.

\subsubsection{Additional units.} While the units mentioned above have covered the most important activities impacted by GenAI in HEIs, there are some less common units that complement AI guidelines in HEIs
. For instance, University Relations (U8) can provide guidance on AI for university marketing and communications, which ``does not address academic use by students or faculty''. The Office of Research and Innovation (U4) focuses on the use of AI in scholarly activity, similar to the role of libraries in other institutions:

\quotes{This document outlines best practices for employing generative AI in various research processes, ensuring its application supports the university's mission while adhering to legal and ethical standards.}{U4}

Additionally, the Office of Student Conduct and Community Standards (U2) and Learning.IU (U6) publishes guidance from a student's perspective, emphasizing the importance of students to proactively seek guidance before using AI:
\quotes{It is your responsibility to know and follow your instructor’s expectations. Expectations will vary across courses. If unsure, check your course syllabi, course information in Canvas, or talk with your instructors.}{U2}

The variety of units involved in AI governance demonstrates the organizational complexity of HEIs and the profound implications of GenAI on higher education. While each unit has its specific emphasis, there is also a significant proportion of overlapping or cross-referencing in their guidelines, which is mainly related to teaching practices or AI usage in the classroom. Although the efforts from multiple units contribute to the comprehensiveness of AI guidelines, it can be difficult for a university member to identify all information relevant to their role from multiple units, creating challenges for guiding responsible AI usage.

\subsection{Role-specific Governance of AI}
The involvement of multiple units reveals another important characteristic of AI governance in HEIs: the need to address the concerns of different roles within the university community. In the case of Big Ten, four predominant roles merged from the AI guidelines: faculty (or instructor), student, staff, and researcher. Some guidelines apply to all members regardless of their roles, such as the data-sharing policy published by IT, but most of the guidelines are written with a specific role as the intended audience, as detailed below. 

\subsubsection{Faculty.} Delivering high-quality education is one of the most important objectives of universities and faculty play an indispensable role in this process. Therefore, it is unsurprising that most guidelines are written from a faculty's perspective:

\quotes{At the TLTC, we look forward to helping you think creatively about your assessments and your specific learning outcomes to put authentic, relevant, student-centered learning at the forefront of your academic planning}{U3}

As this statement suggests, although some AI guidelines are written for faculty, their content is still student-centered, aiming to improve learning outcomes. Specifically, university guidelines often imply three tasks for the role of a faculty. First, the guidelines present plenty of resources for faculty to learn what AI is, what its benefits and limitations are, how it may impact assessment and student learning, how it may impact education equity, etc. Therefore, the first task is for the faculty to become knowledgeable about AI. Although there is no enforcement or specific requirements for what faculty's level of knowledge should be on these topics, the following statement shows universities' preferences for faculty to be more familiarized with AI:

\quotes{AI is quickly becoming an embedded element of the teaching and learning process that requires the acknowledgment and attention of instructors, instructional designers, and academic leaders. }{U13}

Second, faculty are encouraged to integrate AI into their pedagogical practices such as lecture preparation, assignment design, and brainstorming for active learning activities. On the one hand, the need to integrate AI is motivated by the increased productivity brought by AI, e.g., `shortening the time instructors spend on creating course materials, coming up with examples and assignments, as well as making grading more efficient. (U12)' On the other hand, inappropriate use of AI among students can hamper the learning process, making `AI-proof' teaching practices necessary:

\quotes{Probably the best way to guard against inappropriate use of AI-generated text is to redesign your assignments, both the prompts themselves and the related processes.}{U6}

This concern also leads to the third task for faculty: providing guidance for students and supervising students' AI usage. The guidelines support faculty in creating syllabus statements, communicating expectations of AI usage with students, and discussing the implications with students. While instructors have the freedom to define acceptable AI usage in their classes, the guidelines have made recommendations regarding the detection of AI-generated content. Generally, universities discourage the use of AI detection tools, highlighting their high false-positive rates and emphasizing the need to focus on the learning process rather than the assignment product:
\quotes{The available tools are simply not effective in providing the evidence needed to build an academic integrity case against a student. Our pedagogies should be built with critical AI literacy in mind, so it’s important to think through what goals AI prohibition is going to meet and whether enforcement is how you want to spend your time and energy.}{U14}

\subsubsection{Student.} Students typically make up the largest population in the university community and AI guidelines are also student-centered, highlighting the need to promote student learning. However, as the regulation of students' use of AI has been primarily delegated to faculty, AI guidelines are often not written from the student's perspective. The limited guidelines targeting students echo those written for faculty, urging students to communicate with their instructors and seek suggestions and guidance. In addition, the guidelines highlight that students should thoroughly consider the consequences of their usage of AI both for themselves but also for the society:

\quotes{As GenAI poses to be a revolutionary tool that can change the academic space and beyond, it is important for you to understand why and how you intend to use these new, powerful tools... Understand that your usage of GenAI-based tools can give you the means to better not just yourself, but also society as a whole, and there is an ethical responsibility towards doing so.}{U7}

\subsubsection{Staff.} Staff can be involved with a broad range of administrative and operational tasks that support the university community. However, due to the variety of staff positions and their unique responsibilities, it is difficult to provide specific guidance for this role. U8 provided guidance for staff involved in university communication and marketing, demonstrating the importance of these activities for HEIs. In addition to examples of using AI operational tasks, writing, and editing, the guidance also provides examples where GenAI should not be used: 

\quotes{At this time, we advise AI should not be used in the creation of institution-specific content (e.g. leadership messaging) or information regarding the immediate health and safety of our community (e.g. updates and triage.)
}{}
\quotes{Generative AI should not be used to modify any University trademarks, mascots, or otherwise without explicit permission from University Relations.)
}{U8}

These statements show the university's utter caution about how AI may impact the authenticity, accountability, and reliability of its communications.

\subsubsection{Researcher.} While the role of a researcher can often coexist with the other three roles, it indicates the role's involvement with core scholarly activities including conducting research, gathering literature, publishing scholarly work, and more. Since all Big Ten are research-intensive universities, AI usage in research is an important topic in their guidance. Similar to suggestions for faculty, researchers are encouraged to utilize AI too to strengthen productivity in their research processes such as literature review, experiment design, and data analysis. Meanwhile, researchers are advised to take responsibility for their AI usage in research, documenting their usage and accounting for bias and limitations:

\quotes{Researchers should utilize GenAI systems in research only where they perform well and exhibit few hallucinations. Researchers should verify all outputs for accuracy and attribution and attest that this has been done in all cases, detailing the methods used to do so.}{U7}

Overall, these guidelines attempt to address the various concerns of the university community regarding AI usage by providing specific guidelines for four distinct roles. Among these roles, the guidelines for faculty are the most detailed, as faculty play a core role in maintaining educational values and quality while also governing AI usage among students within the classroom context. Guidelines for researchers are also prominent, reflecting the importance of research activities in the Big Ten. In contrast, a smaller portion of the content is dedicated to student-specific or staff-specific perspectives.

\subsection{The Academic Characteristics of AI Governance}
The multi-unit and role-specific characteristics of AI guidelines reflect the organizational complexity and multifaceted functionality of HEIs. However, regardless of the unit and role, AI governance in the Big Ten has generally incorporated academic characteristics, emphasizing the intention to empower the university community for informed decision-making. We summarize the characteristics as \textit{Educative and advisory}, \textit{Flexible}, and the\textit{Socratic method}, as described below:

\subsubsection{Educative and Advisory Guidance.} Except for the IT policies related to data sharing and privacy issues, most of the guidelines do not enforce or prohibit any actions for the university community. Instead, they focus on providing resources for people to learn more about AI and recommend possible behaviors to avoid risks and promote benefits:

\quotes{Purdue University continues to support the autonomy and choice of faculty and instructors to utilize instructional technology that best suits their teaching and learning environments. As such, there is no official university policy restricting or governing the use of Artificial Intelligence, Large Language Models or similar generative technologies.}{U10}

\begin{figure}[h]
\centering
\includegraphics[width=\linewidth]{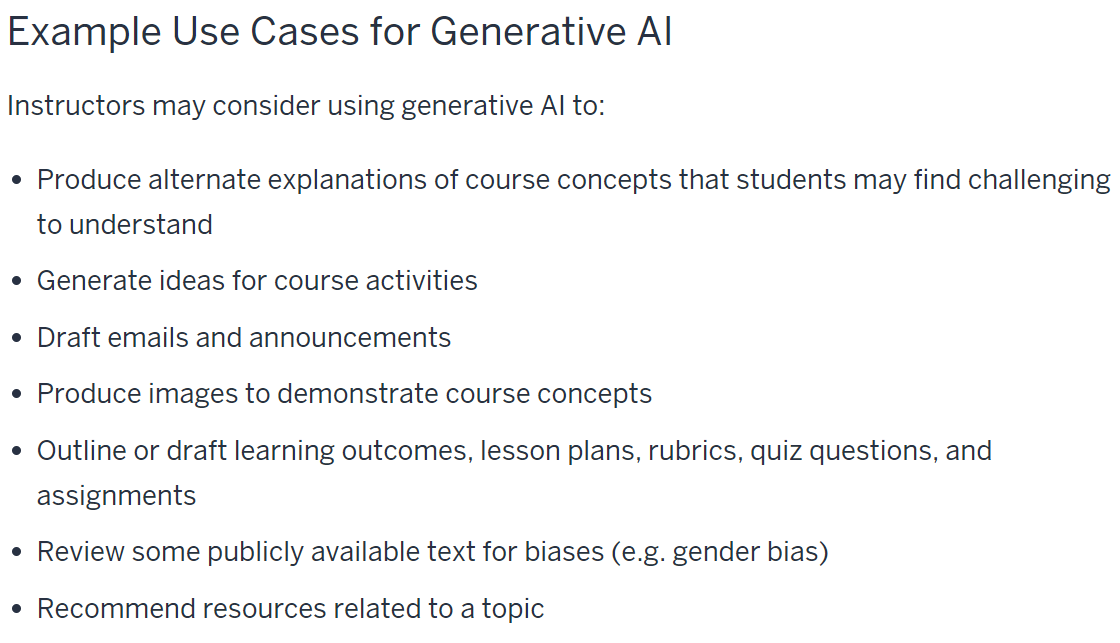}
\caption{Example use cases for GenAI provided by U6}
\label{fig:example}
\end{figure}

As this statement shows, the guidelines are written with the mindset to support the autonomy of its community to make the best use of the technology. In addition to being educative, the guidelines also make specific suggestions on appropriate or inappropriate usage of AI (see Figure~\ref{fig:example}). However, it should be noted with the educative and advisory characteristics, there also comes uncertainty regarding the use and management of AI in some scenarios. An obvious example is the use of AI detection tools. U12 concluded that ``We do not recommend using this detection tool as the basis for reporting a suspected case of academic dishonesty,'' which seems a consensus among the Big Ten. Yet, it remains unclear for faculty regarding what could be used as the basis for reporting academic dishonesty related to AI usage.

\subsubsection{Flexible Guidance.} On the positive side, the uncertainty also accompanies the flexibility in AI guidelines, which encourages the community to experiment and explore the potential benefits of using AI. The flexibility is exemplified through the acknowledgment that AI is a rapidly evolving technology and that the guidelines will evolve as new information arises. In some cases, universities also make the process of developing AI guidelines an engaging and interactive process where different community members are encouraged to share their perspectives on the implications of GenAI. For instance, U14 leveraged the power of social media and created an online space for GenAI discussions among the university community:

\quotes{Welcome to our budding community, a space where we hope to see collaboration and knowledge exchange thrive.  Here, you can both contribute and gain insights into the innovative ways in which our faculty, instructors, students, and staff are using GenAI tools to develop new teaching and learning methodologies. In addition, we hope that this platform will serve as a forum for thoughtful and respectful conversations to address the ethical complexities of GenAI.}{U14}

Moreover, this flexibility is also demonstrated through multiple options provided for resolving AI-related issues. For instance, considering the course syllabus managing students' use of AI, universities often provide three types of sample statements allowing faculty to customize for their classroom setting, e.g., \textit{no restrictions}, \textit{allow limited usage of ChatGPT}, and \textit{prohibit the usage of ChatGPT}.

\begin{figure*}[h]
\centering
\includegraphics[width=0.8\linewidth]{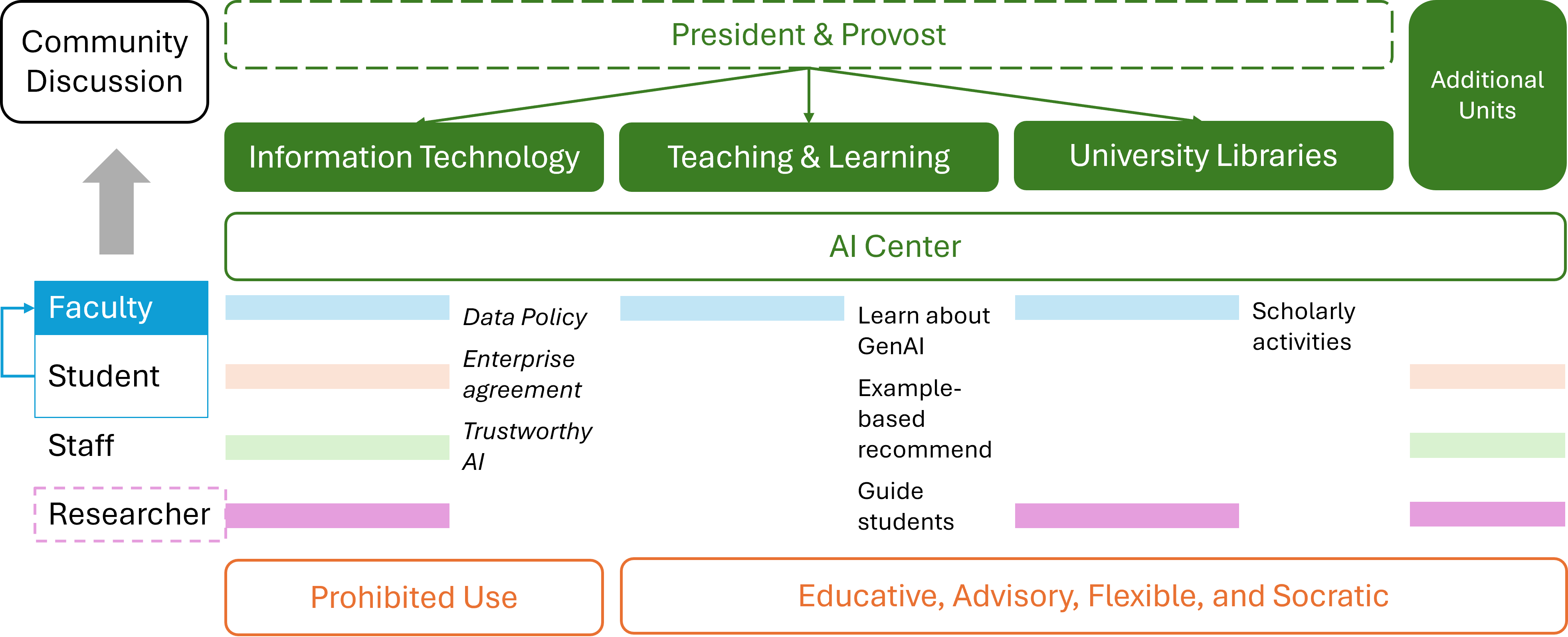}
\caption{Summarized framework of AI Governance through Official Guidance}
\label{fig:framework}
\end{figure*}

\subsubsection{Socratic Method.} The Socratic Method is a form of logical argumentation that promotes critical thinking, provoked by the continual probing questions of the teacher \cite{delic2016socratic}. While the AI guidelines are mostly static descriptions, some of its approaches resemble the Socratic Method by imposing questions on its readers. For some questions, the guidelines will provide possible answers or rationale, creating a FAQ section, but there are also questions that come with no answers, provoking the readers to think critically about the impact of GenAI:

\quotes{As GenAI poses to be a revolutionary tool that can change higher education and beyond, it is important for you to understand why and how you intend to use these new, powerful tools. These are a few questions to consider and note that the answers to these questions will vary for each person.}{U7}

Again, the Socratic Method approach resonates with other academic characteristics, highlighting that the overall intention of AI guidance in universities is to improve AI literacy and raise awareness of responsible AI usage among its community rather than regulating its usage in a strict sense.

\subsection{Summary} The strategies and characteristics of the AI guidelines can be summarized as the framework in Figure~\ref{fig:framework}. The governance of AI is practiced through AI guidelines published by multiple units. Among them, the President \& Provost recapitulates the content from other units and directs people's attention to the resources available. The AI Center can act as the one-stop shop for all guidelines. Other units tend to address the concerns of specific roles in the university community, including faculty, students, staff, and researchers, with guidance for students often delegated to faculty. Regardless of the role, community members are encouraged to participate in the discussion of the GenAI implications. Only the guidelines published by IT specify scenarios of prohibited use of AI. The rest of the guidelines tend to be educative, advisory, flexible, and Socratic, demonstrating the objective to improve the community's understanding of AI and empower them to use GenAI responsibly.

\section{Discussion}

In this section, we discuss the findings in relation to prior literature and their practical implications. First of all, the guidelines published by the Big Ten agree with the literature on the ethical development and deployment of AI: attempting to maximize the benefits of AI but also acknowledging the complex and opaque nature of AI systems and their potential limitations ~\cite{akinrinola2024navigating,diakopoulos2020accountability}. They also demonstrate the recognition of the considerations of human values and rights in the community~\cite{diaz2023connecting}. By contrast, the unique strategies and characteristics of AI guidelines in Big Ten reflect the challenges faced by HEIs when adapting themselves to the advancement of AI technology. 

Specifically, the multi-unit governance of AI helps address the challenge of diverse stakeholders during technology diffusion in HEIs. By engaging different units in the process of AI governance, the university can consider more factors that impact the technology implementation such as technological infrastructure and human resources \cite{ramdhani2021diffusion}. As our findings suggest, each of the units has a specific emphasis on roles and activities impacted by GenAI, contributing to more comprehensive guidelines addressing the needs of different stakeholders. Despite its advantages, the involvement of multiple units can result in more web pages and a more complicated information structure of the guidelines. Consequently, it is inevitable that the level of difficulty in finding information also increases due to the complicated structure \cite{tombros2005users}. For instance, a faculty member may need to access IT, Teaching \& Learning, and University Libraries to find all the information they need, as data privacy, teaching implications, and research use of AI are all related to the role of faculty and they are addressed in different levels of detail by these units. Moreover, as our analysis shows, the content can also overlap between different units, leading to further efforts in comparing those guidelines and checking for discrepancies. Therefore, while the involvement of multiple units is beneficial, the publishing and rendering of AI guidelines may need to be simplified to make it more accessible to the university community.

One possible solution is to leverage the AI Center, i.e., create a website devoted to AI-related issues and make all AI guidelines accessible on this single site. In addition, strategies in website design could used to optimize the guidelines and reduce information load for readers \cite{chen2018improving}. Specifically, information seeking is easier when user needs are considered in the information architecture \cite{shih2012enhancement}, suggesting the guidelines could be designed as role-oriented. Currently, the guidelines provide role-specific content that addresses the concerns of different roles. However, a few of them are organized in a role-specific way, making it difficult for students, faculty, and staff to find corresponding information. Organizing the guidelines in a role-oriented way may help the university community to comprehend the guidelines and put them into practice.

Another important issue regarding different roles is that faculty is the primary focus in most AI guidelines, with the responsibility for guiding students largely delegated to them. While guiding students through classroom-level management may be an effective way to maintain academic quality \cite{korpershoek2016meta} and addressing potential issues brought by AI, it should be noted that this responsibility may lead to an increased workload for faculty. Recent research has shown that faculty could complain about the increased workload due to the need to manage more AI-related academic integrity problems \cite{10.1145/3614419.3644014}. This is especially important as the lack of work-life balance has been a known problem for junior faculty in the United States \cite{azevedo2022examining}. Therefore, it is important for HEIs to consider how they may effectively guide students' use of AI through faculty without creating an extra workload for them.

The workload issue may be further amplified by some of the academic characteristics of AI guidance. While the educative, advisory, and flexible nature gives the community valuable autonomy to make their own decisions about AI usage, it also requires them to become more knowledgeable about AI to make informed decisions. One potential problem is that faculty and students may have difficulty knowing what to do even after reading the guidelines. For example, identifying inappropriate use of AI in coursework is a known struggle for faculty and instructors \cite{10.1145/3614419.3644014} and the flexibility regarding this issue would not make it easier. Similarly, the Socratic Method is ineffective when there is no active participation from students or the question is misaligned with the student's knowledge level. For example, students may not think seriously about the questions listed in the guidelines. Even if they do, it's difficult for them to know exactly how AI could impact the future job market due to their lack of relevant knowledge and perspectives \cite{vicsek2024younger, abdelwahab2023business}. Strengthening the communication between the institution and its community, as well as the community's active participation, can improve engagement and reduce uncertainty in interpreting AI guidelines \cite{moon2023searching}. This communication can include targeted educational sessions, accessible resources, and open forums for discussion, enabling students to build their knowledge base. By fostering a more informed community, students can actively participate in AI governance and take stronger ownership of their AI usage.

It should also be emphasized that the academic characteristics of AI guidance reflect its intention to improve AI literacy and empower the university community to explore the possibilities of AI in a somewhat protected manner. It provides insights for other institutions or even organizations outside the academia in terms of AI governance: it is important to engage the community in defining responsible AI usage and encourage the community to take responsibility for their actions when using AI tools. Other HEIs may utilize the framework in Figure~\ref{fig:framework} to develop their own AI guidelines but also seek ways to balance empowerment and uncertainty in AI governance.

\section{Limitation}
There are limitations in the current study that can be addressed through further research. First, while focusing on the Big Ten allowed us to identify meaningful patterns in AI guidance for HEIs, we recognize that the findings should not be generalized without considering cultural and socioeconomic differences among HEIs across the world. Specifically, HEIs in non-English-speaking countries and those in the Global South may face different challenges at the organizational level, and insights from the Big Ten may not be applicable in their scenarios. Future studies may seek to compare AI governance in universities with different cultural and societal backgrounds. Second, even though most AI guidelines are straightforward and clearly written, readers may still have different perceptions that misalign with the guidelines' intention. Investigating the perception of AI guidelines among different university members is important for assessing the effectiveness of AI guidelines in HEIs' AI governance.

\section{Conclusion}
The current study presents a case study of AI guidance in the Big Ten universities. Through thematic analysis of different AI guidelines, we identify the multi-unit governance of AI, role-specific governance of AI, and the academic characteristics of AI governance in these universities. These strategies and characteristics reflect the university's intention to develop comprehensive guidelines for AI usage that both maintain the autonomy of its community and help them to take advantage of AI. However, the complicated information structure and flexibility in guidance may cause problems and confusion for the community. The findings provide practical implications for other HEIs and organizations regarding AI governance.

\section{Positionality Statement}

The authors acknowledge their backgrounds and the potential biases that might be introduced when interpreting the findings of this study. The primary author's training spans multiple disciplines, including engineering, management, and educational research, and encompasses diverse roles such as student, instructor, and researcher in HEIs in the United States. Similarly, the research team, with its substantial experience in HEIs comparable to those in the Big Ten, brought a comprehensive perspective to the data analysis. All team members actively engaged in interpreting the findings and discussing the implications of the study. Despite this breadth of experience, we recognize that our educational and cultural backgrounds may have shaped our interpretations. To address potential biases, we rigorously documented our initial assumptions and continuously reflected on our understanding at each stage of the research.

\section{Ethic Statement}
All data utilized were publicly available at the time of data collection. It is important to note that the information may have been updated or changed afterward. Furthermore, the interpretations and conclusions drawn represent the authors' viewpoints. While they are based on the universities' AI guidelines, they should not be taken as the official positions or attitudes of the universities. 

\section{Acknowledgment}
The authors would like to thank Jiyoon Kim for contributing to the data collection. The authors also thank Dr. Sarah Zipf, Tehniyet Azam, and Zimeng Shao for their suggestions and help for improving this paper.

\section{Supplementary Materials}
\url{https://github.com/chuhao-w/mindmaps}

\bibliography{aaai24}
\end{document}